\documentclass[]{article}
\usepackage[utf8]{inputenc}
\usepackage{multirow,etoolbox,refcount,multicol,makecell,tikz,verbatim,lmodern,graphicx,microtype,endnotes,titlesec,nameref,hyperref,booktabs,tabularx,comment}
\usepackage[overload]{textcase}
\usepackage[style=nature,autocite=superscript,sorting=none]{biblatex}
\usepackage[separate-uncertainty=true]{siunitx}
\usepackage[justification=justified,labelsep=period,font={small},labelfont={bf},width=0.75\linewidth]{caption}
\usepackage{mathtools,amsthm,upgreek,amssymb}
\usepackage[letterpaper,margin=25mm]{geometry}
\usepackage{filecontents}

\titleformat{\section}[block]
  {\normalfont\fillast}
  {}
  {1ex minus .1ex}
  {\small\bfseries\MakeUppercase}
\titlespacing{\section}
  {2ex}{4ex}{2ex}[2ex]

\titleformat{\subsection}[block]
  {\normalfont}
  {}
  {0ex}
  {\itshape}
\titlespacing{\subsection}
  {0ex}{3ex}{1ex}
  
\newcommand*\deriv[0]{\ensuremath{\frac{d\theta}{dt}}}
\newcommand*\dderiv[0]{\ensuremath{\frac{d^2\theta}{dt^2}}}

\begin{document}

\title{Finding the period of a simple pendulum}
\author{Nicolas Graber-Mitchell}
\maketitle

\begin{abstract}
    Pendulums have long fascinated humans ever since Galileo theorized that they are isochronic with regards to their swing.
    While this simplification is useful in the case of small-angle pendulums due to the accuracy of the small-angle approximation, it breaks down for large-angle pendulums and can cause larger problems with the computational modelling of simple pendulums.
    This paper will examine the differences between the periods of small-angle and large-angle pendulums, offering derivations of the period in both models from the basic laws of nature.
    This paper also provides a common way of deriving elliptic integrals from physical phenomena, and the period of pendulums has been one of the major building blocks in this new, developing field.
    Lastly, this paper makes a number of suggestions for extensions into the study of simple pendulums that can be performed.
    While this paper is not intended as a rigorous mathematical proof, it is designed to illuminate the derivation of the exact periods of simple pendulums and carefully walks through the mathematics involved.
\end{abstract}

\tableofcontents

\section{Introduction} \label{sec:intro}
Simple pendulums consist solely of a rod fixed at one end and attached to a weight at the other.
When released, they swing.
As one of the first to study pendulums, Galileo famously determined that their motion is independent of the swept angle.
While observing a chandelier, he saw that simple pendulums are \textit{isochronic} with regards to the maximum angular displacement: no matter the angle a given pendulum makes with the vertical in its swing, its period of motion will always be the same.\autocite{helden_pendulum_1995}
If this is true, then pendulums exhibit simple harmonic motion, which occurs when the force that pushes a system to equilibrium is proportional to the displacement from that equilibrium.\autocite{nave_simple_2005}
Since Galileo's observation, pendulums have led breakthroughs in time-measurement and science in general, at one point proving that the Earth rotates once a day.

In high school physics courses, we learn to treat pendulums as isochronic.
While modelling their motion, we follow Galileo's assumption and ignore the swept angle while calculating the period of a pendulum.
In fact, we find these results for ourselves in earlier years of school.
I can remember timing the swings of washers tied to strings in fifth grade, eventually coming to the same conclusion as Galileo.
Surprisingly, Galileo actually was wrong: the maximum angular displacement of the pendulum does affect its oscillatory period.
While pendulums with relatively small maximum displacements are, for all intents and purposes, isochronic, the error associated with this assumption increases with the angle.

The purpose of this investigation is to analyze how Galileo's assumption is both right and wrong, and in what cases.
By deriving two differential equations from the basic laws of nature that govern pendulums, one representing the approximation and one representing the exact period, and conducting a comparison, the error cases of the approximation can be found.
This is extremely important: since a pendulum's motion is periodic, any slight difference in the time of the period when modelling can result in huge errors after a few oscillations.
After solving these differential equations, we can also investigate why the small-angle approximation is used.
Does it really make a difference in the difficulty of the problem?
If the exact solution for period is easy to use, then it should be used instead of the approximation.
However, if the approximation is not terribly inaccurate and is much easier to employ, then it makes sense to do so.
Most of all, however, this investigation sets out to explore a common piece of physics knowledge that we all learn at a fairly early age, and to determine how wrong it really is.

\begin{figure}[htb]
    \centering
    \scalebox{0.8}{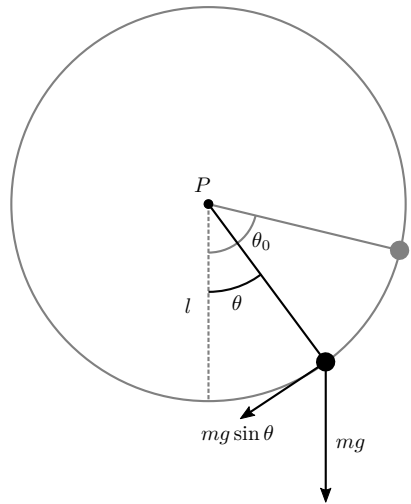}
    \caption{Diagram of a simple pendulum.}
    \label{fig:pendulum}
\end{figure}

\begin{figure}[htb]
    \centering
    \scalebox{0.7}{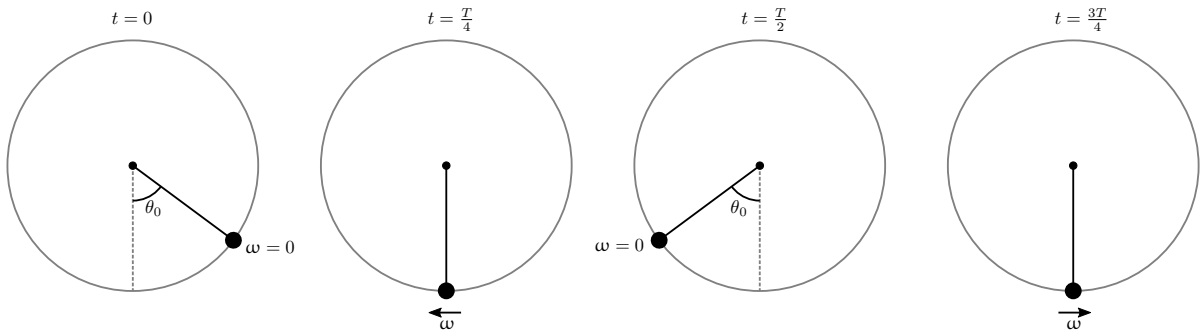}
    \caption{Periodic oscillating motion of a released pendulum. \(\upomega\) denotes the angular velocity, and so \(\upomega = \deriv\).}
    \label{fig:osc}
\end{figure}

\section{Modeling simple pendulums} \label{sec:model}

\subsection{Definition of a simple pendulum}

A simple pendulum consists of a mass \(m\) hanging from a rigid, massless rod of length \(l\) fixed at a point \(P\) in a vertical plane, as shown in Figure \ref{fig:pendulum}.
A force of gravity with acceleration \(g\) straight downwards can act on the mass.
The pendulum is free to swing, and when released from a maximum angular displacement \(\theta_0\) it oscillates, swinging back and forth from \(\theta = \theta_0\) to \(\theta = -\theta_0\).
This motion is demonstrated in Figure \ref{fig:osc}, and is periodic: one full oscillation takes a time of length \(T\) to return back to the original position.

\subsection{Application of Newton's laws}

The pendulum is governed by Newton's laws applied to rotational motion:
\begin{equation}
    \tau = I \alpha , \label{eqn:laws}
\end{equation}
where \(\tau\) is torque, \(I\) is moment of inertia, and \(\alpha\) is the angular acceleration.\autocite{daniel_acoustics_2011,boston_university_physics_department_torque_1999}
Torque is a measure of rotational force---a twist.
The moment of inertia of an object, also known as its rotational inertia, is a measure of how difficult it is to change the rotation of that object.
It is an analogue of mass: mass is to force as rotational inertia is to torque.
Angular acceleration is simply the change in velocity of the angle \(\theta\) of the pendulum.

A torque on a lever can be expressed as \(\tau = F r\), a force \(F\) multiplied by a distance from the pivot \(r\).
In this case, \(r = l\) and \(F = mg \sin \theta\), the force due to gravity that is perpendicular to the string.
This force can be found by splitting the total gravitational force into its parallel and perpendicular parts through vector algebra.
This operation is common and physics.
In this case, it yields \(F = mg \sin \theta\): the perpendicular force is the product of the total force \(mg\) and the scale factor \(\sin \theta\).

Moving on to the right hand side of (\ref{eqn:laws}), we need to define the moment of inertia of the pendulum.
The moment of inertia is found in any given object by \( I = \sum mr^2\) of each point in the object.
The mass in our pendulum is concentrated in a single point at the bob, and so the summation has a single term: this point.
Therefore, \(I = ml^2\).

Combining these new equivalent expressions for the torque and the moment of inertia,
\begin{align}
    -l m g \sin \theta &= m l^2 \alpha , \\
\intertext{then simplifying and rearranging,}
    -\frac{g}{l} \sin \theta &= \alpha .
\end{align}
Notice that the torque was made negative.
This is by convention, since a negative torque indicates force in the clockwise direction, and the force of gravity perpendicular to the pendulum rod is clockwise at the initial state of the pendulum.

For simplicity, let \(w = \sqrt{\cfrac{g}{l}}\).
Next, as angular acceleration is the second derivative with respect to time of angular position, \(\alpha = \dderiv\).
Simplifying, we obtain the differential equation that governs the motion of simple pendulums:
\begin{equation} \label{eqn:diff}
    \dderiv = -w^2 \sin \theta 
\end{equation}

Thus, the motion of the pendulum depends solely on the length \(l\) of the string, the acceleration \(g\) due to gravity, and the angle \(\theta\) that the pendulum makes with the downward-facing vertical vector.

\section{Small-angle pendulums}
For pendulums with small \(\theta_0\), \(\sin \theta \approx \theta\), which is known as the small-angle approximation, hence the name ``small-angle pendulum.''
In order to find the approximation, we use the squeeze theorem, which says \autocite{austin_christian_math_2016}
\begin{equation}
    \lim_{x \to 0} \frac{\sin x}{x} = 1 .
\end{equation}
Therefore, when \(x \to 0, \, \sin x = x\) and so then \(\sin x \approx x\).
This approximation can be used to simplify the differential equation in (\ref{eqn:diff}), yielding
\begin{equation}
     \dderiv \approx -w^2 \theta . \label{eqn:approxdiff}
\end{equation}
The equation (\ref{eqn:approxdiff}) is a much simpler second order differential equation than (\ref{eqn:diff}), and its solution is used to find the period of a simple pendulum in many high school physics classes, including my own last year.

In order to solve this differential equation, we first notice that that the second derivative \(\dderiv\) is simply the function itself scaled by the constant \(-w^2\).
This form is known as a second order linear homogeneous differential equation.\autocite{tseng_second_2016}
The parent function that expresses such a recursive behavior is the exponential equation: \(\frac{d}{dx} \, e^x = e^x\).
Letting \(\theta = e^{iwt}\), we find the first and second derivatives:
\begin{align}
    \theta &= e^{iwt} \\
    \deriv &= iwe^{iwt} \\
    \dderiv &= -w^2 e^{iwt} = -w^2 \theta .
\end{align}

Therefore, \(\theta_1 = C_1 e^{iwt}\) is a solution to the second order differential equation in (\ref{eqn:approxdiff}), where \(C_1 \in \mathbb{C}\).
Notice also that \(\theta_2 = C_2 e^{-iwt}\) is a solution, where \(C_2 \in \mathbb{C}\).
The constant coefficients \(C_1\) and \(C_2\) are preserved through differentiation and are essential integration constants.
These constants repesent the infinite number of solutions to the differential equation in (\ref{eqn:approxdiff}) determined by the initial characteristics of the pendulum.
The sum of these solutions is a more general solution, since either of the two constants be be set to \(0\) to obtain the original solutions:
\begin{align}
    \theta  &= C_1 e^{iwt} + C_2 e^{-iwt} . \\
\intertext{This can be shown by taking the derivative once,}
    \deriv  &= C_1 iw e^{iwt} - C_2 iw e^{-iwt} , \\
\intertext{and then differentiating again to obtain}
    \dderiv &= - C_1 w^2 e^{iwt} + C_2 w^2 e^{-iwt} \nonumber \\
    &= -w^2 \left( C_1 e^{iwt} + C_2 e^{-iwt} \right) \nonumber \\
    &= -w^2 \theta . \label{eqn:almost}
\end{align}
Therefore, (\ref{eqn:almost}) satisfies (\ref{eqn:approxdiff}) and encompasses all of the possible solutions to the problem.
 
Applying Euler's formula \(e^{ix} = \cos x + i \sin x\) to this general solution, we obtain
\begin{equation}
    \theta = C_1 \cos wt + C_1 i \sin wt + C_2 \cos \! \left( -wt \right) + C_2 i \sin \! \left( -wt \right) .
\end{equation}
As \(\sin \! \left( -\theta \right) = -\sin \theta\) and \(\cos \! \left( -\theta \right) = \cos \theta\),
\begin{equation}
    \theta = C_1 \cos wt + C_1 i \sin wt + C_2 \cos wt - C_2 i \sin wt .
\end{equation}
After rearranging,
\begin{equation}
    \theta = \left( C_1 + C_2 \right) \cos wt + i \left( C_1 - C_2 \right) \sin wt ,  
\end{equation}
and then letting \(A = C_1 + C_2\) and \(B = i \left( C_1 - C_2 \right)\),
\begin{equation}
    \theta = A \cos wt + B \sin wt .
\end{equation}
In this equation, \(A, B \in \mathbb{R}\) are constants defining the initial conditions of the pendulum.
Since \(\theta\) is a linear combination of periodic functions, it is also periodic.
The periods of both \(\cos wt\) and \(\sin wt\) are \(\frac{2\pi}{w}\).
Adding two periodic functions with the same period results in another periodic function with that same period, and therefore the period of a pendulum after applying the small-angle approximation is
\begin{equation}
    T \approx \frac{2\pi}{w} = 2\pi \sqrt{\frac{l}{q}}
\end{equation}

Notice that this approximation of period has no reliance on the maximum angular displacement of the pendulum.
This is in accordance with Galileo's observations laid out in the \textsc{\nameref{sec:intro}}.

\section{Large-angle pendulums}
However, for pendulums with large \(\theta_0\), the small-angle approximation is no longer applicable.
In order to obtain the period of such a pendulum---known as large-angle---we must solve the differential equation (\ref{eqn:diff}) without the use of the approximation.

In order to integrate the second order differential equation, notice that
\begin{equation} \label{eqn:funky}
    \frac{d}{dt} \left[ \frac{1}{2} \left( \deriv \right)^2 \, \right] = \dderiv \deriv
\end{equation}
by the chain rule. Therefore, taking the antiderivative of the right hand side, effectively reversing the operation:
\begin{equation}
    \int \dderiv \deriv \, dt = \frac{1}{2} \left( \deriv \right)^2 + C.
\end{equation}
In order to make this antiderivative applicable to our problem, we multiply the original differential equation (\ref{eqn:diff}) by \(\deriv\),
\begin{align}
    \dderiv \deriv &= -w^2 \sin \theta \, \deriv , \\
\intertext{and the left hand side becomes the product present in (\ref{eqn:funky}). Now integrating with respect to \(t\),}
    \int \dderiv \deriv dt &= \int -w^2 \sin \theta \, \deriv dt , \\
\intertext{the differential becomes}
    \frac{1}{2} \left( \deriv \right)^2 &= w^2 \cos \theta + C, \\
\intertext{where \(C\) is the constant of integration. Next, as \(\deriv = 0\) when \(\theta = \theta_0\), we can set \(C = - w^2 \cos \theta_0\) to obtain}
    \frac{1}{2} \left( \deriv \right)^2 &= w^2 \cos \theta - w^2 \cos \theta_0 . \\
\intertext{Rearranging and solving for \(\deriv\), we obtain}
    \deriv &= w \sqrt{2} \sqrt{\cos \theta - \cos \theta_0}. \label{eqn:cos}
\end{align}
At this point, our original second order differential equation has been solved into a first order differential equation.
However, this is still not enough to make the period of the pendulum apparent.

Before we continue, it is important to note a few identities.
By the sine half-angle identity,
\begin{equation}
    \sin \frac{\theta}{2} = \pm\sqrt{\frac{1 - \cos \theta}{2}}.
\end{equation}
Solving for cosine, we get a modified version of the cosine double-angle identity:
\begin{equation} \label{eqn:double}
    \cos \theta = 1 - 2\sin ^2 \frac{\theta}{2}.
\end{equation}

Applying the identity in (\ref{eqn:double}) to (\ref{eqn:cos}), \(\cos \theta = 1 - 2\sin ^2 \frac{\theta}{2}\) and \(\cos \theta_0 = 1 - 2\sin ^2 \frac{\theta_0}{2}\).
After substitution and simplification, 
\begin{align}
    \deriv &= 2w \sqrt{\sin ^2 \frac{\theta_0}{2} - \sin ^2 \frac{\theta}{2}}. \\ 
\intertext{Let \(k = \sin \frac{\theta_0}{2}\), which is a constant, and substitute to obtain}
    \deriv &= 2w \sqrt{k^2 - \sin ^2 \frac{\theta}{2}}. \label{eqn:sine} \\
\intertext{Next, we take the reciprocal of (\ref{eqn:sine}) in order to integrate with respect to \(\theta\):}
    \frac{dt}{d\theta} &= \frac{1}{2w \sqrt{k^2 - \sin ^2 \frac{\theta}{2}}} . \label{eqn:inv}
\end{align}

We integrate from \(0\) to \(\frac{T}{4}\) on the left hand side and from \(0\) to \(\theta_0\) on the right hand side, since as the pendulum sweeps out an arc from the bottom of its circle to its maximum angular displacement, a time equal to \(\frac{T}{4}\) passes.
Therefore, these integrals are equivalent.
Integrating (\ref{eqn:inv}), we obtain
\begin{align}
    \int_0^\frac{T}{4} \frac{dt}{d\theta} \: d\theta &= \int_0^{\theta_0} \frac{d\theta}{2w \sqrt{k^2 - \sin ^2 \frac{\theta}{2}}}, \\
\intertext{and completing the integration on the left hand side then simplifying and moving constants out of the integral,}
    T &= \frac{2}{w} \int_0^{\theta_0} \frac{d\theta}{\sqrt{k^2 - \sin ^2 \frac{\theta}{2}}}. \label{eqn:period}
\end{align}
Thus, we have the period of a simple pendulum obtained from the differential equation (\ref{eqn:diff}).
This period depends on the length \(l\) of the rod, the acceleration \(g\) due to gravity, and the maximum angular displacement \(\theta_0\) of the pendulum.

Curiously, the integral in (\ref{eqn:period}) is known as an elliptic integral of the first kind, and is extremely important in the field of cryptography.
The inverses of elliptic integrals are known as elliptic functions, and they have numerous applications beyond that of pendulums.
Elliptic integrals are proven to lack explicit solutions and must be integrated numerically.

\begin{figure}[hp]
\centering
\input{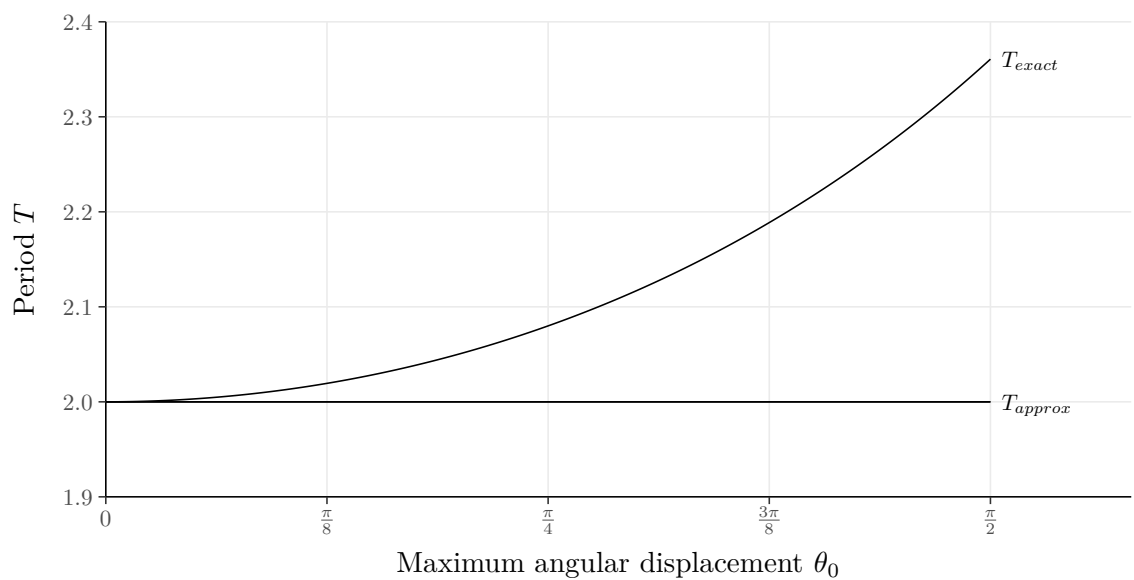}
\captionof{figure}{The exact and approximate periods of a pendulum defined by \(w = \pi\).}
\label{plot:direct}
\end{figure}
\begin{figure}[hptb]
\centering
\input{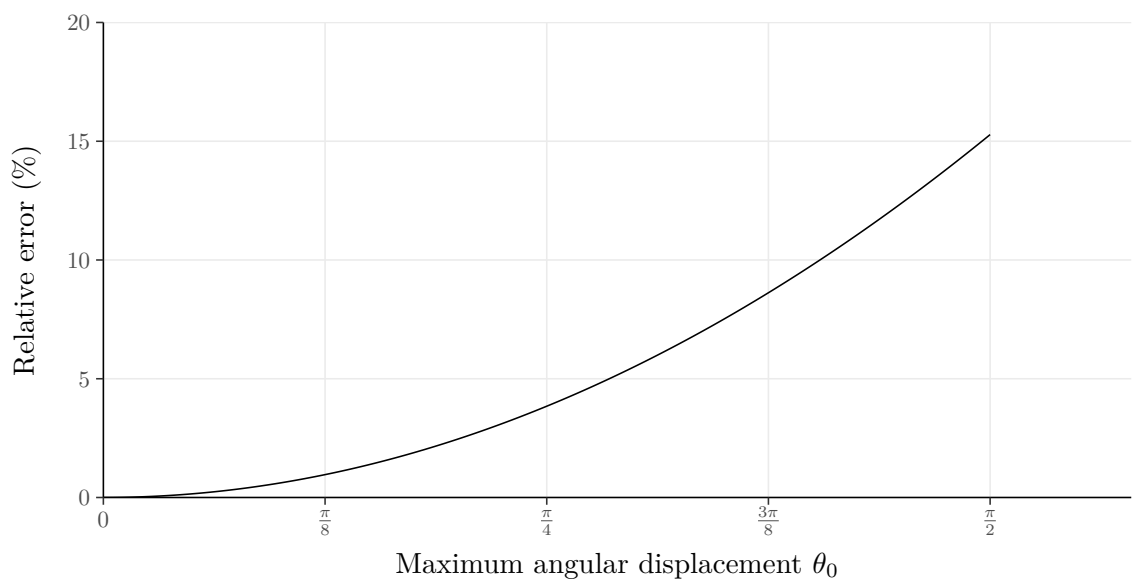}
\captionof{figure}{The error of the small-angle approximation as related to the maximum angular displacement.}
\label{plot:error}
\end{figure}

\section{Evaluating the error of the small-angle approximation}

In the previous two sections, we found two expressions for the period of a simple pendulum: one assuming the small-angle approximation
\begin{alignat}{2}
    T(g, l) &\approx T_{approx} &&= \frac{2\pi}{w} \label{eqn:approx} \\ 
\intertext{and one employing a definite integral (see \textsc{\nameref{sec:con}} for discussion on this integral)}
    T(g, l, \theta_0) &= T_{exact} &&= \frac{2}{w} \int_0^{\theta_0} \! \frac{d\theta}{\sqrt{k^2 - \sin ^2 \frac{\theta}{2}}} . \label{eqn:exact}
\end{alignat}
Both (\ref{eqn:approx}) and (\ref{eqn:exact}) can be numerically evaluated provided constants outlined in \textsc{\nameref{sec:model}}.
While both are functions of the acceleration due to gravity \(g\) and the length of the pendulum \(l\), only the exact period is a function of the maximum angular displacement \(\theta_0\).

Next, we can identify how far off the approximation is at larger values of \(\theta_0\).
Using the equation for percent relative error
\begin{equation}
    R = 100 \, \frac{T_{exact} - T_{approx}}{T_{exact}},
\end{equation}
we obtain the graph shown in Figure \ref{plot:error}.

\begin{table}[h!]
\renewcommand{\arraystretch}{1.3}
\centering\small
\begin{tabular}{@{}cllr@{}}\toprule
\(\theta_0\) & \(T_{approx}\) & \(T_{exact}\) & R (\%) \\ \midrule
0 & 2 & --- & --- \\
\(\frac{\pi}{8}\) & 2 & 2.0194 & 0.96 \\
\(\frac{\pi}{4}\) & 2 & 2.0799 & 3.84 \\
\(\frac{3\pi}{8}\) & 2 & 2.1887 & 8.62 \\
\(\frac{\pi}{2}\) & 2 & 2.3607 & 15.28 \\
\bottomrule
\end{tabular}
\centering
\captionof{table}{The periods associated with different maximum displacements in a period with \(w = \pi\).}
\label{table:period} 
\end{table}

We can also evaluate both \(T_{approx}\) and \(T_{exact}\) at a variety of points and calculate the error, yielding Table \ref{table:period}.
In both Table \ref{table:period} and Figure \ref{plot:direct}, the constant \(w\) that takes into account the gravitational force and length of the pendulum is set to \(\pi\).
The \(\pi\) term then drops out of the approximation, making comparison easier.

\section{Conclusion} \label{sec:con}
The small-angle approximation in pendulum period is fairly good.
When \(\theta < \frac{\pi}{4}, \, R < 3.84\%\).
However, even this small inaccuracy can add up: after about 13 full oscillations of a pendulum with \(w = \pi\) and \(\theta_0 = \frac{\pi}{4}\), the approximation is out of phase by half a period with the exact pendulum.
This means that after 13 oscillations, when the approximate pendulum is at equilibrium, the exact pendulum is maximally displaced, and vice-versa.
It's clear to see that these errors matter when conducting longer models of the motion of the pendulum, but on a short time scale, the error is almost negligible---especially with small angles, such as \(\frac{\pi}{8}\), where the approximation is already less than a percent inaccurate.
Pendulums with small angles such as these are not uncommon: many clocks have long, thin pendulums, with a high \(l\) and a low \(\theta_0\).
These two constants lend themselves to producing a long, accurate period.

As expected, the small-angle approximation gets worse as \(\theta_0\) grows.
Since the approximation relies on the property that \(\lim_{x \to 0} \frac{\sin x}{x} = 1 \), it makes sense that it will become less accurate as \(x \to \infty^+\).
Important to note is that the approximation becomes less accurate faster near higher values of \(\theta_0\).
While at low values the error \(R\) is both low and increasing slowly, at high values of \(R\) the error is high and increasingly quickly.
For all \(\theta_0 \in \, \, ] \, 0, \pi \, [\), \(\deriv > 0\) and \(\dderiv > 0\).

Lastly, the small-angle approximation is a definite improvement in usability over the exact equation.
It is impossible to obtain an algebraic expression for the definite integral present in (\ref{eqn:exact}), which is known as an elliptic integral.
First discovered in the search for the length of an ellipse's curve and the period of a pendulum, they were expanded upon and are now relevant to many branches of mathematics.
While \(T_{exact}\) contains an elliptic integral that can be evaluated only numerically, \(T_{approx}\) does not, and is in fact extremely simple.
It makes sense to use the approximation whenever possible.
Even if the maximum angular displacement of a pendulum is relatively large, it makes more sense to use the approximation when not worried about minutiae.
However, in an accurate computer model, the exact equation should be used.
Otherwise, the phase errors discussed earlier will greatly influence the motion of the pendulum over time.

\subsection{Extensions}
This investigation has a number of extremely interesting and useful extensions that should be followed up on.
While many of them relate to theoretical math alone, experimental data-gathering would also be useful.

First, it would be valuable to corroborate the truthfulness of the exact equation (\ref{eqn:exact}) through experimental results.
Using a laser tripwire attached to a computer and a pendulum, the period of the pendulum's swing with varied maximum angular displacement can be compared to the curve found in Figure \ref{plot:direct}.
While a certain amount of error would be introduced into the data by the impossibility of some of the characteristics of the ideal pendulum described in \textsc{\nameref{sec:model}} (such as the inability to construct a pendulum with a rod of zero mass), the results should correspond to the relationship defined by \(T_{exact}\).

Secondly, the exact equation might be viable as a method of approximating \(\pi\).
Interestingly, both \(T_{exact}\) and \(T_{approx}\) are products of \(\frac{2}{w}\) and another factor.
In the approximation, that factor is \(\pi\)---a constant.
In \(T_{exact}\), that factor is an integral reliant on the maximum angular displacement.
Since the approximation becomes more and more correct as this displacement gets closer to 0, the result of this definite integral must get closer and closer to \(\pi\) as the displacement gets closer to 0.
More formally,
\begin{equation}
    \lim_{\theta_0 \to 0} \, \int_0^{\theta_0} \! \frac{d\theta}{\sqrt{k^2 - \sin ^2 \frac{\theta}{2}}} = \pi .
\end{equation}
It's fascinating how \(\pi\) can show up in unexpected places, given that this integral does not necessarily require the use of \(\pi\) to numerically evaluate or even obtain in the first place.
In fact, without using the small-angle approximation, \(\pi\) does not show up in this problem at all except for as the limit of this integral.

Lastly, elliptic integrals are incredibly important in everyday life, driving new, modern cryptographic applications.
When you access your bank account, there is a good chance that your actions are being secured by elliptic curve cryptography.
An extension of this project would be to derive common elliptic functions from the elliptic integral in (\ref{eqn:exact}), and then discuss their usefulness and properties.
Such a pursuit would mirror the original discoveries of these integrals and their transition from applications derivable from real-life phenomena to fields not found in nature.

\nocite{*}
\printbibliography
\addcontentsline{toc}{section}{References}

\end{document}